\documentclass[twoside,onecolumn,english,aps,showpacs,prx,superscriptaddress]{revtex4-1}
\usepackage[T1]{fontenc}
\usepackage{color}
\usepackage{bm}
\usepackage{amsmath}
\usepackage{amsthm}
\usepackage{amssymb}
\usepackage{graphicx}
\usepackage{esint}
\usepackage{mathrsfs}
\usepackage{booktabs}
\usepackage{multirow}
\usepackage{makecell}
\usepackage{verbatim}
\usepackage[plainpages = false, pdfpagelabels, bookmarksnumbered = true,
                 breaklinks = true,
                 linktocpage,
                 colorlinks = true,
                 linkcolor = blue,
                 urlcolor  = blue,
                 citecolor = blue,
                 anchorcolor = green,
                 hyperindex = true,
                 hyperfigures
                 ]{hyperref} 
\usepackage{dcolumn}  
\DeclareMathOperator{\Tr}{Tr}

\newtheorem{theorem}{Theorem}



\begin{document}

\preprint{APS/123-QED}

\title{Detecting entanglement with transport measurement in weakly interacting and fluctuating systems}
\author{Zhenhua Zhu}
\affiliation{State Key Laboratory of Low Dimensional Quantum Physics, Department of Physics, Tsinghua University, Beijing, 100084, China}
\affiliation{Frontier Science Center for Quantum Information, Beijing 100184, China}
\author{Gu Zhang}
\affiliation{National Laboratory of Solid State Microstructures, School of Physics, Jiangsu Physical Science Research Center and Collaborative Innovation Center of Advanced Microstructures, Nanjing University, Nanjing 210093, China}
\author{Dong E. Liu}
\email{Corresponding to: dongeliu@mail.tsinghua.edu.cn}
\affiliation{State Key Laboratory of Low Dimensional Quantum Physics, Department of Physics, Tsinghua University, Beijing, 100084, China}
\affiliation{Frontier Science Center for Quantum Information, Beijing 100184, China}
\affiliation{Beijing Academy of Quantum Information Sciences, Beijing 100193, China}
\affiliation{Hefei National Laboratory, Hefei 230088, China}

\begin{abstract}
    Measuring entanglement entropy in interacting, multipartite systems remains a significant experimental challenge.
    We address this challenge by developing a protocol to measure von Neumann entropy (VNE) and mutual information in quantum transport systems with both many-body interactions and multiple subsystems.
    Our analysis indicates that the vital connection between VNE and two-point correlation functions persists under these realistic conditions.
    The measurement is shown to be feasible for systems with boundary interactions and, critically, for bulk-interacting systems subject to a quantum quench of their internal couplings.
    Our work provides a pathway to experimentally quantify entanglement in complex interacting systems and establishes mutual information as an experimentally accessible indicator for system-environment entanglement.
\end{abstract}


\begingroup
\let\clearpage\relax   
\let\cleardoublepage\relax
\maketitle
\endgroup

\section{Introduction}
Quantum entanglement serves as a cornerstone of quantum technologies. Its principles are crucial to a wide array of research fields, from quantum computing~\cite{jozsa2000distinguishability,kak2007quantum,adami1997neumann,zeng2015quantum,eisert2006entanglement} to the characterization of complex phenomena like quantum phase transitions~\cite{osborne2002entanglement,yang2005reexamination,osterloh2002scaling} and quenches~\cite{mitra2018quantum,alba2017entanglement,coser2014entanglement,alba2018entanglement}.
A key focus within this domain is the indicator of quantum entanglement, for which a variety of theoretical tools, e.g., Bell inequality~\cite{werner2001bell,bartkiewicz2013entanglement,barr2024quantum}, entanglement witness~\cite{lewenstein2000optimization,chruscinski2014entanglement,eisert2007quantitative}, and the Peres-Horodecki Criterion~\cite{peres1996separability,simon2000peres}, have been introduced.
One of the most versatile of these is the mutual information, defined via the von Neumann entropy (VNE).
It has been successfully applied to probe many-body localization~\cite{de2017quantum}, establish bounds on connected correlation functions~\cite{wolf2008area}, explore signatures of quantum chaos~\cite{hosur2016chaos,seshadri2018tripartite}, and signify the entanglement between a system and its environment~\cite{hosur2016chaos,wanisch2021delocalization}.

Despite the fundamental importance of VNE and mutual information, their direct experimental measurement remains a significant challenge. A key protocol, first introduced in Ref.~\cite{klich2009quantum}, is especially constrained to non-interacting bipartite systems. To date, measurement protocols have been largely restricted to systems of confined quasi-particles~\cite{islam2015measuring,pichler2016measurement,lewis2019dynamics}, with recent extensions to certain many-body quantum transport systems~\cite{klich2009quantum,song2012bipartite,ZhangNC24,gullans2019entanglement}. These limitations hinder its application to the interacting, multipartite systems that are often of experimental interest.

In this work, we develop a generalized protocol to probe VNE and mutual information experimentally in quantum transport, transcending bipartite non-interacting paradigms~\cite{klich2009quantum,song2012bipartite,ZhangNC24,gullans2019entanglement}. Our theory encompasses two key contributions: (i) multipartite systems, beyond the bipartite case, and (ii) the effects of interactions, in contrast to non-interacting models.
Our central result is that the relation between VNE and two-point correlation functions holds even in these more general interacting, multipartite settings [cf. Eqs.~\eqref{result}, \eqref{generating}, and Theorems \ref{tm1}, \ref{tm2}].
We further show that VNE is experimentally accessible in systems where interactions are localized at the boundaries (e.g. at and/or close to the island, see Fig.\ref{fig.1}).
For systems with bulk (e.g. the leads, see Fig.\ref{fig.1}) interactions, we find that the entropy and mutual information remain measurable when the coupling between subsystems is abruptly switched between the fully transparent and fully disconnected limits.
Our work substantially broadens the applicability of entanglement quantification frameworks beyond previous constraints~\cite{klich2009quantum,song2012bipartite}. This advancement enables entanglement detection in more realistic quantum systems, characterized by multi-partite structures and interactions, while also furnishing a concrete indicator of system-environment entanglement.

\section{Multi-partite entanglement}
For an isolated bipartite system, its VNE quantifies the quantum entanglement between two composing subsystems.
More generally, for a multi-partite isolated system, addressing entanglement between its subsystems involves another quantity, i.e., the so-called mutual information.

For a multi-partite system $W = \{R_1,R_2\ldots R_x\}$, its 
$x$-partite mutual information at time $t$ is defined from VNE:
\begin{equation}\label{MI}
\begin{split}
    I^{(x)} (t, R_1:\cdots: R_x) = 
    \sum_{\mathcal{R}\subseteqq W} (-1)^{N(\mathcal{R})-1} S_{(\mathcal{R})},
\end{split}
\end{equation}
where $S_{(\mathcal{R})} = \operatorname{Tr}_\mathcal{R}(\rho_\mathcal{R} \log \rho_\mathcal{R})$ is the VNE of the subset $\mathcal{R}$, containing $N(\mathcal{R})$ subsystem(s) of $W$, with $\rho_\mathcal{R}$ its reduced density matrix.
Following its definition, Eq.~\eqref{MI}, mutual information quantifies the total correlation—both classical and quantum—shared between subsystems. It is determined by the VNEs of the individual subsystems and their joint state, thereby capturing the total shared information that cannot be accounted for by examining the subsystems in isolation~\cite{zeng2015quantum}.
To capture the dynamical feature of the mutual information, we further define
\begin{equation}
    \Delta I^{(x)}(t_0, t_0 +\Delta t) \equiv I^{(x)}(t_0 +\Delta t) - I^{(x)}(t_0)
    \label{eq:mutual_information_production}
\end{equation}
as the change of the mutual formation during the time interval $\Delta t$.
In Eq.~\eqref{eq:mutual_information_production} the labels of subsystems have been neglected for simplicity.

Critically, one of the key applications of the paper is establishing multipartite mutual information as an environmental coupling criterion in quantum devices.
For an $x$-partite system $W$, its mutual information can potentially serve as an effective criterion for identifying whether $W$ is isolated, i.e., when it entirely decouples from external environments.
Indeed, with an odd $x$, $\Delta I^{(x)}$ of $W$ is zero if $W$ is isolated, and becomes generically finite otherwise~\cite{hosur2016chaos,wanisch2021delocalization}.
A finite $\Delta I^{(x)}$ can thus signify the system-environment coupling, after partiting the system into an odd number of subsystems.
Notice that this criteria however fails if $x$ is instead an even number.

\section{Detection of VNE and mutual information}
Following Eq.~\eqref{MI}, detecting mutual information production $\Delta I^{(x)}$ of a system requires a protocol to measure VNE production, $\Delta S_{(\mathcal{R})}$ [cf. Eq.~\eqref{result}] of any subset $\mathcal{R}$, which in-turn 
requires a generalization of Refs.~\cite{klich2009quantum,song2012bipartite} (on VNE of bipartite systems) to multi-partite systems.
Corresponding analysis involves two-point correlation functions, defined as $M_{(\mathcal{R})}^{ij} \equiv \Tr_\mathcal{R} \rho_\mathcal{R} c_i^\dagger c_j$, where $\Tr_\mathcal{R}$ is the partial trace over the reduced density matrix $\rho_\mathcal{R}$ of the subset $\mathcal{R}$, and $c_i$ annihilates one state of the subset $\mathcal{R}$.

Briefly, for a multi-partite system, we find that the VNE production of any subset $\mathcal{R}$ (that contains one or multiple subsystems) can be obtained via (see details in \ref{supple})
\begin{equation}\label{result}
\begin{split}
\Delta S_{(\mathcal{R})}(t_i, t_f)&=\sum_{m>0}\frac{\alpha_m}{m!}C_m^{(\mathcal{R}\bar{\mathcal{R}})}(t_i, t_f),\\
\alpha_m&=\begin{cases}
(2\pi)^m|B_m|,& \text{m even}\\
0,& \text{m odd}
\end{cases},
\end{split}
\end{equation}
where $\bar{\mathcal{R}}$ refers to the complement of $\mathcal{R}$ (with respect to the system $W$), $\Delta S_{(\mathcal{R})}(t_i,t_f)$ represents the VNE produced between the initial ($t_i$) and final moments ($t_f$).
$B_m$ denotes Bernoulli numbers
and $C_m^{(\mathcal{R}\bar{\mathcal{R}})}$ represents the $m$-th cumulant (in the perspective of full counting statistics) between subsets $\mathcal{R}$ and $\bar{\mathcal{R}}$.
For instance, $C_1^{(\mathcal{R}\bar{\mathcal{R}})}$ and $C_2^{(\mathcal{R}\bar{\mathcal{R}})}$ refer to the average number of tunneling electrons between $\mathcal{R}$ and $\bar{\mathcal{R}}$, and the corresponding electron number fluctuation, respectively.
These cumulants can be obtained via the
generating function:
\begin{align}
&\chi(\lambda,\mathcal{R},\bar{\mathcal{R}}):=\sum_{N_{\mathcal{R}\bar{\mathcal{R}}}}P(N_{\mathcal{R}\bar{\mathcal{R}}})e^{i\lambda N_{\mathcal{R}\bar{\mathcal{R}}}}\nonumber\\
&C_m^{(\mathcal{R}\bar{\mathcal{R}})}=\frac{\partial^m}{\partial (i\lambda)^m}\log[\chi(\lambda,\mathcal{R},\bar{\mathcal{R}})],
\label{generating}
\end{align}
that describes particle transport between subsets $\mathcal{R}$ and $\bar{\mathcal{R}}$~\cite{levitov1993charge,klich2009quantum,song2012bipartite}.
Here $\lambda$ is the ``counting field'', and
$P(N_{\mathcal{R}\bar{\mathcal{R}}})$ is the probability to transmit $N_{\mathcal{R}\bar{\mathcal{R}}}$
particles between subsets $\mathcal{R}$ and $\bar{\mathcal{R}}$, during the interval $t_i$ and $t_f$. 

Importantly, although Eqs.~\eqref{result} and \eqref{generating} are in-principle bipartite (i.e., dealing with entanglement between a subset $\mathcal{R}$, and the rest of the system), they can be used to obtain the mutual information [defined by Eqs~\eqref{MI} and \eqref{eq:mutual_information_production}] of multi-partite systems, after summing over necessary system subset options.

\begin{figure}[t]
\centering
\includegraphics[width=0.48\textwidth]{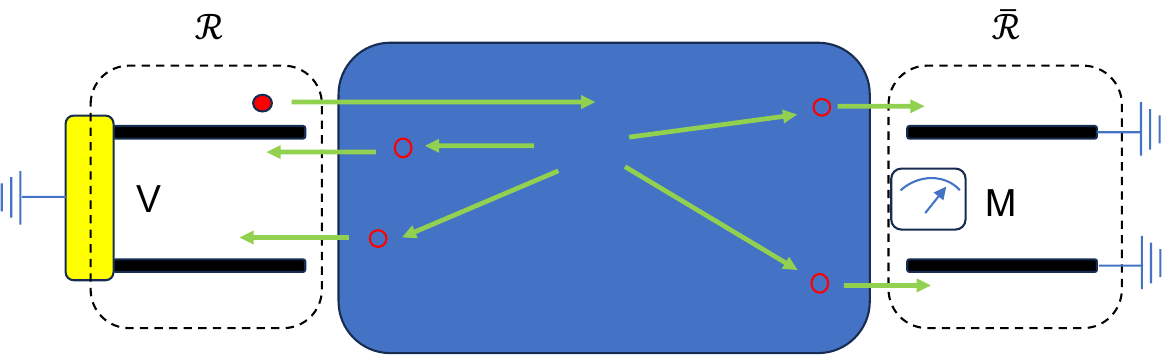}
\caption{A sketch of the measurement setup for multi-terminal island systems. The island (blue), whose charging energy is sufficiently large compared to the tunneling strength, is coupled to the leads (black). By applying a voltage bias (yellow) to a subset $\mathcal{R}$, a wave package will (represented by a solid red circle) travel across the island and scattering to each leads (hollow red circles). The green lines highlight the scattering directions. A monitor $M$ is used to simultaneously detect the current signal in the channels between leads in $\bar{\mathcal{R}}$ and island, which provides $C_m^{\mathcal{R}\bar{\mathcal{R}}}$.}
\label{fig.1}
\end{figure}

It is important to emphasize that our scheme captures the non-trivial physical origin of entanglement generated by quantum scattering. As illustrated in Fig.~\ref{fig.1}, this non-local entanglement arises from the splitting of electron wave packets as they scatter across the central mesoscopic quantum system (the island). Although decoherence rapidly sets in as the electron wave packets leave the island and propagate into the leads, this decoherence obscures the direct detection of quantum information through conventional transport measurements. However, our detection scheme is sensitive to the accumulated entanglement, as shown in Eq.~\eqref{result}. Therefore, even though transport measurements are inherently macroscopic, they still provide a means of probing the quantum properties of microscopic scattering processes within the island, highlighting the subtle and non-trivial nature of quantum phenomenon.
Eqs.~\eqref{eq:mutual_information_production} to \eqref{result} further provide the experimental accessibility to mutual information in multi-partite systems, and whether the system couples to an outer environment.
Remarkably, our work specifically address only the quantum devices and their transport signatures as shown in Fig.~\ref{fig.1}, complementary to those studying the many-body bulk systems [e.g.,~\cite{berkovits2015entanglement,lim2024mesoscopic,froland2024entanglement}].

\section{Weakly interacting case}
The relation between two-point correlation functions and von Neumann entropy (VNE), crucial for entanglement detection in transport, is well-understood in non-interacting fermionic Gaussian states (FGS)~\cite{peschel2003calculation}. However, as interactions are prevalent in experiments~\cite{kouwenhoven2001revival,beri2012topological,esslinger2010fermi}, this section therefore addresses this relation in the weakly interacting regime. Remarkably, we find the fundamental connection persists in this regime, as stated in the following theorem:
\begin{theorem}\label{tm1}
Consider a fermionic system with conserved charges, where the effective Hamiltonian includes interaction terms
whose interaction strength $\Lambda_{ijkl}$ are sufficiently weak in density matrix so that we can apply perturbation theory. For example, $\beta|\Lambda_{ijkl}|\ll1$ is required in thermal equilibrium states where $\beta=1/k_BT$ with $k_B$ the Boltzmann constant and $T$ the temperature. By partitioning the system into two subsets $\mathcal{R}$ and $\bar{\mathcal{R}}$, the von Neumann entropy between subset $\mathcal{R}$ and the remaining subset $\bar{\mathcal{R}}$ is, up to first order in the interaction strength $\Lambda$, given by a standard form:
\begin{align}\label{theroem_1}
S_{(\mathcal{R})}=&-\Tr[(1-M_{(\mathcal{R})})\log(1-M_{(\mathcal{R})})+M_{(\mathcal{R})}\log{M_{(\mathcal{R})}}],
\end{align}
where the correlation matrix of subset $\mathcal{R}$ is defined as $M_{(\mathcal{R})}^{ij}=\Tr_\mathcal{R} \rho_\mathcal{R} c_i^\dag c_j$,
with $\rho_\mathcal{R}$ the reduced density matrix of subset $\mathcal{R}$, and $c_i^\dag$, $c_j$ the fermionic creation and annihilation operators of single-particle states in subset $\mathcal{R}$.
\end{theorem}

Proving this theorem simply involves an expansion of $\rho_{\mathcal{R}}$ in weakly interacting systems around the FGS, up to the first order of interaction.
For instance, without the loss of generality, we can assume a thermal state influenced by a quartic interaction $H_t=\sum_{ijkl}\Lambda_{ijkl}\tilde{c}_i^\dag \tilde{c}_j\tilde{c}_k^\dag \tilde{c}_l$ 
where $\tilde{c}_i$ denotes the original operator (distinct from $c_i$), and $c_i$ (defined in Theorem \ref{tm1}) are the fermionic operators obtained by diagonalizing the quadratic part of the reduced density matrix.
At leading order, the reduced density matrix then becomes
\begin{equation}\label{rhoeff}
\rho_\mathcal{R}=\frac{e^{-\sum_is_i'n_i}}{Z_{\mathcal{R}}} \Big[1+\Big(\sum_{ijkl}\Lambda_{ijkl}g_{ijkl}\Big)^l+\mathcal{O}(\Lambda^{l+1}) \Big].
\end{equation}
Here $n_i=c_i^\dag c_i$, $s_i'n_i$ denotes the effective action without interaction and $g$ is an operator consisting of $c^\dag$ and $c$.
The second term contains the leading order contribution from the weak interaction, to an integer order $l\geq 1$, which depends on the explicit expression of $H_t$.
With the expression of $\rho_\mathcal{R}$, one can obtain Eq.~\eqref{theroem_1} by treating $\log\rho_\mathcal{R}$ as an operator and dealing with its expectations (see \ref{II}). We emphasize that our proof relies on the validity of perturbation theory, thereby justifying retention of solely leading-order interaction contributions.
Note that Eq.~\eqref{theroem_1} does not imply the absence of corrections from interactions. Actually, it is implicitly contained by two-point correlation functions, i.e., elements of $M_{(\mathcal{R})}$.
Indeed, following its definition, $M_{(\mathcal{R})}^{ij}=\Tr_\mathcal{R} \rho_\mathcal{R} c_i^\dag c_j$, the interaction-involved matrix elements are generically different from their interaction-free counterparts.

Remarkably, Eq.~\eqref{rhoeff} can be achieved with a general non-equilibrium scenario as [see~\ref{V}]:
\begin{equation}\label{noneq}
\rho = e^{-i(H_2+H_4)t}\rho_0e^{i(H_2+H_4)t}/Z,
\end{equation}
where $\rho_0$ denote an initial thermal state with weak interaction and $H_2~(H_4)$ is the quadratic (quartic) Hamiltonian. We consider the short $t$ and weak $H_4$ regime so that $iH_4t$ perturbs the action. As a consequence, Theorem \ref{tm1} holds beyond thermal states [cf.~\ref{V}].

Theorem \ref{tm1} demonstrates that even in weakly interacting cases, the VNE can be approximately described using only two-point correlation functions.  Crucially, the spectral density of these functions is measurable within certain interaction regimes, a point elaborated upon later. Consequently, Eq.~\eqref{theroem_1} facilitates the detection of VNE and mutual information in weakly interacting systems. 

It is crucial to elucidate how our approach circumvents the replica trick \cite{calabrese2009entanglement}. The replica trick calculates VNE using $S_{(\mathcal{R})} = -(\partial_n\Tr_\mathcal{R}{\rho_\mathcal{R}^n})|_{n=1}$ to bypass direct calculations of $-\Tr_\mathcal{R}{\rho_\mathcal{R} \log{\rho_\mathcal{R}}}$. In our method, weak interactions simplify the density matrix to an exponential form akin to FGS. This allows direct expression of $\log{\rho_\mathcal{R}}$ using creation and annihilation operators, facilitating simpler VNE calculations.

\section{Weak fluctuation limit around the saddle point}
Previous analysis demonstrated that, in the presence of weak interactions, the entanglement-correlation relation, Eq.~\eqref{theroem_1}, retains the structural form of the non-interacting case. The influence of the interaction is implicitly captured within the correlation matrix $M_{(\mathcal{R})}$. Furthermore, it is shown that this conclusion remains valid beyond the perturbative limit, extending to the weak fluctuation regime.

More specifically, as shown by e.g., Ref.~\cite{coleman2015introduction}, a fermionic quartic interaction can be removed, at the expense of introducing an auxiliary bosonic field $\varphi$. After tracing out the fermionic degrees of freedom (with the Hubbard-Stratonovich transformation), the effective partition function $Z_{eff}$
now contains quartic and higher order interactions defined with bosonic operators (see more details in \ref{III}).
To remove the bosonic interaction, it is standard to introduce the saddle point that is defined as $\partial_\varphi Z_{eff}|_{\varphi_0}=0$.
Especially, when the saddle point is stable, i.e., $\partial_\varphi^2Z_{eff}|_{\varphi_0}>0$, fluctuations of the bosonic modes, i.e., $\propto (\varphi-\varphi_0)^2$ become small and irrelevant, such that we are allowed to take the saddle-point approximation.
With this approximation, we recover a result analogous to the weakly interacting case, as formalized in the following theorem:

\begin{theorem}\label{tm2}
We consider a charge-conserved fermionic system
described by an effective Hamiltonian with quartic interactions.
Around a stable saddle point, up to the Gaussian fluctuations,
the von Neumann entropy between a subset $\mathcal{R}$ and its complement, $\bar{\mathcal{R}}$, is given by the standard form as shown in Eq.~\eqref{theroem_1}.
\end{theorem}

As outlined above, near the saddle point, the effective partition function becomes quadratic for both fermionic and bosonic modes.
Actually, taking a system in thermal equilibrium as an example, after the Hubbard-Stratonovich transformation, the system density matrix becomes
(see \ref{III}):
\begin{equation}
\begin{split}
\rho=&\frac{1}{Z}e^{-\beta(\sum_{ij}\epsilon_{ij}\tilde{c}_i^\dag \tilde{c}_j-\sum_{ijkl}\lambda_{ijkl}\tilde{c}_i^\dag \tilde{c}_j\tilde{c}_k^\dag \tilde{c}_l)}\\
=&\frac{1}{Z'}\int D[{\phi}]e^{-\beta[\sum_{ij}\epsilon'_{ij}\tilde{c}_i^\dag \tilde{c}_j-\sum_{ijkl}(-\lambda_{ijkl}\phi_{ij}\phi_{kl})]}\\
&\times \left[1+\sum_{ijkl}\phi_{ij}\phi_{kl}G_{ijkl}]\right],
\end{split}
\end{equation}
where $\phi=\varphi-\varphi_0$ represents fluctuations around saddle point, $\epsilon'_{ij}=\epsilon_{ij}-\sum_{kl}(\lambda_{ijkl}+\lambda_{klij})\varphi_{0,ij}$, and $G_{ijkl}$ represents quartic fermionic correction terms. After integrating out the bosonic field and the Hilbert space of region $\bar{\mathcal{R}}$, the reduced density matrix $\rho_\mathcal{R}$ around the saddle point becomes approximately (see \ref{III}):
\begin{equation}\label{rhoeffmf}
\rho_\mathcal{R}=\frac{1}{Z_{\mathcal{R}}}\left[ e^{-\beta\sum_i\tilde{\epsilon}_i'n_i}\left(1+\sum_{ijkl}G'_{ijkl} \right) \right],
\end{equation}
where $G'$
arises from the Gaussian fluctuation around the saddle point.
Theorem~\ref{tm2} can then be achieved by evaluating the expectation value of $\log\rho_\mathcal{R}$ (see \ref{III}). 
Remarkably, Eq.~\eqref{rhoeffmf} also holds in non-equilibrium systems with density matrix in the form of Eq.~\eqref{noneq} [cf.~\ref{V}].

We stress that, in Theorem~\ref{tm1}, the quartic interaction is required to be a small quantity.
Theorem~\ref{tm2}, in contrast, imposes no restriction on the interaction strength. Instead, the system is required to stay in the weak fluctuation limit, i.e., close enough to a stable saddle point.

\section{Entanglement measurement in interacting systems}
Theorems~\ref{tm1} and~\ref{tm2} indicate that for a system within either the weak-interaction or weak-fluctuation limit, its VNE and mutual information are experimentally accessible by measuring the two-point correlation functions $M_{(\mathcal{R})}^{ii}$.
To connect our theory to experimental observables, we further analyze the detection of $M_{(\mathcal{R})}$ for weakly interacting or fluctuating systems.

As the starting point, generically the generating function equals
\begin{equation}\label{generating_int}
\begin{split}
\chi(\lambda)&=\langle e^{i\lambda(\sum_{i\in\mathcal{H}_\mathcal{R}}n_i-N_{\mathcal{R}}^0)}\rangle\\
&=\Tr \left[\rho_{\mathcal{R}} e^{i\lambda(\sum_{i\in\mathcal{H}_\mathcal{R}}n_i-N_{\mathcal{R}}^0)}\right],
\end{split}
\end{equation}
where $\mathcal{H}_\mathcal{R}$ and $N_{\mathcal{R}}^0 $ refer to the Hilbert space, and initial particle number, respectively, of the subset $\mathcal{R}$.
Notice that Eq.~\eqref{generating_int} is generically valid in interaction systems, as its exponential factor, $\sum_{i\in\mathcal{H}_\mathcal{R}}n_i-N_{\mathcal{R}}^0$, equals the total number of particles transferred between subsystems during the measuring time, with or without the presence of interaction.
As shown in Eqs~\eqref{rhoeff} and~\eqref{rhoeffmf}, the reduced density matrix has the form $\rho_{\mathcal{R}}=e^{-\sum_{i\in \mathcal{H}_{\mathcal{R}}}s_in_i}(1-f)/Z_\mathcal{R}$ with $f$ the correction term arising from weak interaction or fluctuation [cf.~\ref{II},~\ref{III} and~\ref{V}]. As a result, the generating function can be rewritten as:
\begin{equation}\label{chi_general}
\log\chi(\lambda)=\log\chi_0(\lambda)-\langle f\rangle_\lambda+\langle f\rangle_{\lambda=0},
\end{equation}
where $\log\chi_0(\lambda)$ accounts for the non-interacting sector and is given by $\sum_i\log(z-M_{(\mathcal{R}),0}^{ii})$ after neglecting irrelevant terms, with $M_{(\mathcal{R}),0}^{ii}$ being the interaction-free two-point correlation function.
Specifically, the irrelevant terms are proportional to $\log z$ and $\log(z-1)$, whose contributions vanish when handling VNE [cf.~\ref{I} and~\ref{IV}].
In Eq.~\eqref{chi_general}, we have defined $\langle f\rangle_\lambda:=\Tr(\rho_\lambda f)$ and $\rho_\lambda:=e^{-\sum_{t\in\mathcal{H}_\mathcal{R}}(s_t-i\lambda)c_t^\dag c_t}/Z_\lambda$, for simplicity.

More specifically, in the remainder of this section we apply Eq.~\eqref{generating_int} and assess the validity of Eq.~\eqref{result} for interacting systems in two distinct regimes: \textit{(i) boundary interactions}, which arise in and immediately around the central island where electron scattering is concentrated (see Fig.~\ref{fig.1}); and \textit{(ii) bulk interactions}, which prevail throughout the extended regions of the tunneling leads.

With interaction present at boundaries, the correction term $f$ is quadratic and the generating function Eq.~\eqref{chi_general} reduces to [cf. \ref{IV} for details]:
\begin{equation}\label{gene-edge}
\log\chi(z)=\sum_i\log(z-M_{(\mathcal{R})}^{ii}),
\end{equation}
where $z=(1-e^{i\lambda})^{-1}$.
With Eq.~\eqref{gene-edge}, the spectral density of matrix $M_{(\mathcal{R})}$, which is essential for obtaining VNE [see Eq.~\eqref{theroem_1} and \ref{I}], can be achieved through $\mu(z):=\Tr\delta(z-M_{(\mathcal{R})})=\frac{1}{\pi}\Im\partial_z\chi(z-i0)$. As a consequence, the VNE (and mutual information) production is related to full counting statistics and Eq.~\eqref{result} holds in the presence of an interaction at the boundaries.

The aforementioned conclusion remarkably applies to a wide range of finite-size boundary interactions, including the exchange coupling in Kondo systems and local electron-electron and electron-phonon interactions~\cite{RosenowHalperinPRL02,PapaMacDonaldPRL04,CarregaSassettiPRL11,SnizhkoCheianovPRB15}.
Actually, outside of the interacting area, particles are asymptotically free.
The transport of these asymptotically free particles through the finite interacting region can be fully characterized by a scattering matrix ($S$-matrix), which treats the interacting area as a ``black box''.
Integrating out the internal degrees of freedom of this region maps the problem onto an effectively non-interacting system, from which the result in Eq.~\eqref{gene-edge} follows.

Analysis above however shows some deviations in bulk-interacting systems, where a finite-size interacting area can not be defined.
For simplicity, here we focus on a quartic in single-particle operators: an assumption that is reasonable in most practical systems (see e.g., Refs.~\cite{song2012bipartite,tam2022topological}).
With interactions in the bulk and supposing the reduced density matrix $\rho_{\mathcal{R}}=e^{-\sum_{i}s_ic_i^\dag c_i}(1-\sum_{k\leq m,l\leq n,m,n}\Gamma_{klmn}c_k^\dag c_l c_m^\dag c_n)/Z_\mathcal{R}$,
the generating function, Eq.~\eqref{generating_int} becomes
\begin{align}
\log\chi(z)&=\sum_i\log(z-M_{(\mathcal{R})}^{ii})\nonumber\\
&-\sum_{k<l,l}\frac{\Gamma_{kkll} M_{(\mathcal{R}),0}^{kk} M_{(\mathcal{R}),0}^{ll} \bar{M}_{(\mathcal{R}),0}^{kk}\bar{M}_{(\mathcal{R}),0}^{ll}}{(z- M_{(\mathcal{R}),0}^{kk})(z- M_{(\mathcal{R}),0}^{ll})},
\label{gene-bulk}
\end{align}
where
$\bar{M}_{(\mathcal{R}),0}^{kk}\equiv 1- M_{(\mathcal{R}),0}^{kk}$.
Importantly, due to the second line of Eq.~\eqref{gene-bulk}, for systems with interaction in the bulk, the relation between entanglement and FCS cumulants, i.e., Eq.~\eqref{result}, requires modifications to the leading-order of interaction.
Nevertheless, Eq.~\eqref{result} remains practically useful, 
when for all $k$, the occupation operator $M_{(\mathcal{R}),0}^{kk}$ equals zero or one. With a zero initial temperature, this is exactly the case where the tunneling between two subsystems abruptly switches between the transparent and the off limits, a situation discussed by Ref.~\cite{klich2009quantum}.
From this perspective, our result indicates that the entanglement entropy measurement protocol of Ref.~\cite{klich2009quantum}, obtained for non-interacting systems, actually remains valid to the leading order of an interaction in the bulk.

Briefly, in systems with interaction in the boundary, the VNE can be experimentally obtained by measuring the cumulants, following Eq.~\eqref{result}.
For systems with bulk interactions, however, the measurement of VNE imposes additional constraints as follows. In the thermal equilibrium case, the lead temperature must be sufficiently low. In the non-equilibrium case [see Eq.~\eqref{noneq}], two conditions must be satisfied:
1) The initial temperature must be sufficiently small; 2) the transmission must be either fully transparent or vanishingly weak. These requirements ensure the applicability of the method under different interaction regimes.

\begin{acknowledgments}
    This work is supported by the Innovation Program for Quantum Science and Technology (Grant No.~2021ZD0302400).
\end{acknowledgments}

\section{Appendix}\label{supple}

\subsection{Multi-partite Entanglement}\label{I}

In this section, we provide details on deriving the relation between entanglement and cumulants of full counting statistics (FCS) [i.e., Eq.~\eqref{result} of the main text], for multipartite fermionic systems.

Here we begin with the FCS generating function $\chi (\boldsymbol{\lambda})$ in multi-partite systems~\cite{klich2009quantum}:
\begin{equation}\label{generating}
\chi(\boldsymbol{\lambda}):=\sum_{\boldsymbol{N}}P(\boldsymbol{N})e^{i\sum_{(p,q)}\lambda_{pq} N_{pq}},
\end{equation}
where $\boldsymbol{N}$ is a matrix, with its element $N_{pq}$ representing the number of particles transferred from subsystems $R_p$ to $R_q$, during the measuring period. Here the measuring period is defined as the interval of time during which the corresponding measuring field $\lambda_{pq}$ of the matrix $\boldsymbol{\lambda}$ is finite.

The evolution of a charge-conserving multi-partite system $W = \{R_1,R_2\ldots R_x\}$ can be described by the scattering matrix
\begin{equation}
\textbf{U}=
\begin{pmatrix}
U_{11}& U_{12}& U_{13}\\
U_{21}& U_{22}& U_{23}& \cdots\\
U_{31}& U_{32}& U_{33}\\
 &\cdots & &\cdots\\
\end{pmatrix},
\end{equation}
where the matrix element $U_{pq}$ refers to the single-particle transmission amplitude from subsystem $R_p$ to subsystem $R_q$. The S-matrix $\textbf{U}$ should be unitary, i.e., $\textbf{U}^{\dag}\textbf{U}=I$.
After turning on the tunneling (represented by the matrix $\textbf{U}$) for a time interval $t$, occupation operator before turning on inter-system transmission, $\textbf{n}$, evolves into $\textbf{n}_U=\textbf{U}\textbf{n}\textbf{U}^\dag$.

Next, we relate the generating function $\chi$ to the occupation operation. As shown by Ref.~\cite{klich2009quantum}, the generating function in non-interacting systems can be written as (the interacting case is discussed in Sec.~\ref{IV}):
\begin{equation}\label{generate}
\chi(\boldsymbol{\lambda})=\det[1-\textbf{n}+\textbf{n}\textbf{U}^\dag \tilde{\textbf{U}}],
\end{equation}
where
\begin{equation}
\tilde{U}_{pq}=e^{i\lambda_{pq}}U_{pq},
\end{equation}
which equals the product between the scattering matrix element, and the corresponding auxiliary field $\lambda_{pq}$.
Alternatively, by splitting the auxiliary field as $\lambda_{pq}=\lambda_p-\lambda_q$, it can be written into the matrix form,
\begin{equation}
\begin{split}
\tilde{\textbf{U}}=&\sum_{p,q}(e^{i\lambda_{pq}}\textbf{P}_p\textbf{U}\textbf{P}_q)
=\sum_{p,q}e^{i\lambda_p}\textbf{P}_p\textbf{U}e^{-i\lambda_q}\textbf{P}_q
=(\sum_{p}e^{i\lambda_p}\textbf{P}_p)\textbf{U}(\sum_{q}e^{-i\lambda_q}\textbf{P}_q)\\
=&(\sum_pe^{i\lambda_p\textbf{P}_p}-(x-1)I)\textbf{U}(\sum_qe^{-i\lambda_q\textbf{P}_q}-(x-1)I)\\
=&e^{i\sum_p\lambda_p\textbf{P}_p}\textbf{U}e^{-i\sum_q\lambda_q\textbf{P}_q},
\end{split}
\end{equation}
where $x$ is the number of subsystems and $\textbf{P}_p$ denotes the projection on the modes of subsystem $R_p$.

As stated in the main text, we can divide the system $W$ into two subsets, $\mathcal{R}$ and $\bar{\mathcal{R}}$, each containing one or multiple subsystems. Following similar steps of Ref.~\cite{klich2009quantum}, by setting $\lambda_p=\lambda_q$ for $\forall R_p, R_q\subset \mathcal{R}$ and $\lambda_j=\lambda_k$ for $\forall R_j, R_k\subset \bar{\mathcal{R}}$, we can write the generating function as:
\begin{equation}
\begin{split}
\chi=&\det\left[z-\left(\sum_{p,R_p\subset\mathcal{R}}\textbf{P}_p\right)\textbf{n}_U\left(\sum_{p, R_p\subset\mathcal{R}}\textbf{P}_p\right)\frac{X}{z}\right],
\end{split}
\end{equation}
where $z=(1-e^{i\lambda_{pj}})^{-1}$ with $R_p\subset\mathcal{R}$, $R_j\subset\bar{\mathcal{R}}$ and $X=e^{-i\textbf{U}[(\sum_{(p,R_p\subset\mathcal{R})}\lambda_{pj}\textbf{P}_p)\textbf{n}]\textbf{U}^\dag}$ with $R_j\subset\bar{\mathcal{R}}$.

Since $M_{(\mathcal{R})}=(\sum_{(p,R_p\subset\mathcal{R})}\textbf{P}_p)\textbf{n}_U(\sum_{(p,R_p\subset\mathcal{R})}\textbf{P}_j)$, the spectral density of $M_{(\mathcal{R})}$, $\mu=\Tr\delta[z-M_{(\mathcal{R})}]$, is
\begin{equation}\label{mu}
\mu(z)=\frac{1}{\pi}\Im\partial_{z}\log\chi(z-i0)+A\delta(z)+B\delta(z-1),
\end{equation}
where $A$ and $B$ are coefficients relating to $X$ and $\dim M_{\mathcal{R}}$. In Eq.~\eqref{mu}, the two delta functions vanish unless $z=1$ or $z=1$, which indicates there irrelevance for von Neumann entropy [as we will show later in Eq.~\eqref{integ}]. Therefore, we will drop them in the following discussion~\cite{klich2009quantum}. 

To relate the generating function to entanglement, we refer to the von Neumann entropy between subsets $\mathcal{R}$ and $\bar{\mathcal{R}}$:
\begin{equation}
S_{(\mathcal{R})}=-\Tr\rho_{\mathcal{R}}\log\rho_{\mathcal{R}},
\end{equation}
where $\rho_{\mathcal{R}}=\Tr_{j,R_j\subset\bar{\mathcal{R}}}\rho$ and $\rho$ is the density matrix of the whole system. While the von Neumann entropy can be written as $S_{(\mathcal{R})}=-\Tr[M_{(\mathcal{R})}\log M_{(\mathcal{R})}+(1-M_{(\mathcal{R})})\log(1-M_{(\mathcal{R})})]$, we can easily rewrite it with the spectral density of $M_{(\mathcal{R})}$:
\begin{equation}\label{integ}
S_{(\mathcal{R})}=-\int_0^1\mu(z)[z\log{z}+(1-z)\log{(1-z})]dz.
\end{equation}
As we just argued, the Integrand in Eq.~\eqref{integ} vanishes at $z=0$ and $z=1$, which results in the irrelevance of the two delta functions in Eq.~\eqref{mu}.  Following the similar steps as in Ref.~\cite{klich2009quantum}, the relationship between quantum noise $C_m$ and von Neumann entropy production is in the form of
\begin{equation}\label{supp_result}
\begin{split}
\Delta S_{(\mathcal{R})}(t_i, t_f)&=\sum_{m>0}\frac{\alpha_m}{m!}C_m^{(\mathcal{R}\bar{\mathcal{R}})}(t_i, t_f),\\
\alpha_m&=\begin{cases}
(2\pi)^m|B_m|,& \text{m even}\\
0,& \text{m odd}
\end{cases},
\end{split}
\end{equation}
where $\Delta S_{(\mathcal{R})}(t_i,t_f)$ represents the VNE produced between the initial ($t_i$) and final time ($t_f$). $B_m$ denotes Bernoulli numbers and $C_m^{(\mathcal{R}\bar{\mathcal{R}})}$ represents the $m$-th cumulant between subsets $\mathcal{R}$ and $\bar{\mathcal{R}}$.

\subsection{Perturbation Theory in Weakly Interacting Systems}\label{II}
In non-interacting case, VNE may be rewritten with two-point correlation function $M_{(\mathcal{R})}$\cite{peschel2003calculation}. A nature question arises: What about interacting case? In this section, we will derive the expression of VNE in weakly interacting systems where the interaction can be treated as perturbation. Remarkably, to calculate VNE with interaction, we will treat $\log\rho_{\mathcal{R}}$ as an operator.

We start with a simple example where the reduced density matrix is in the form of $\rho_{\mathcal{R}}=e^{-S}/Z_{\mathcal{R}}=e^{-S_0-S_I}/Z_{\mathcal{R}}$ with $S_0$ the quadratic part and $S_I$ the weak interaction term. For example, for a thermal equilibrium state with Hamiltonian $H=\sum_{ij}\epsilon_{ij}\tilde{c}_i^\dag \tilde{c}_j+\sum_{ijkl}\Lambda_{ijkl}\tilde{c}_i^\dag \tilde{c}_j \tilde{c}_k^\dag \tilde{c}_l$, $S_0=\beta\sum_{ij}\epsilon_{ij}\tilde{c}_i^\dag \tilde{c}_j$ and $S_I=\beta\sum_{ijkl}\Lambda_{ijkl}\tilde{c}_i^\dag \tilde{c}_j \tilde{c}_k^\dag \tilde{c}_l$ with $\beta =1/k_BT$. Generally, we can always apply a unitary transformation to diagonalize the quadratic terms so that $S_0=\sum_is_{i}n_i$ and $S_I=\sum_{i\leq k,j\leq l,k,l}\lambda_{ijkl}c_i^\dag c_j c_k^\dag c_l$, where $c$, $c^\dag$ denote the new fermionic operators after diagonalizing $S_0$ and $n_i=c_i^\dag c_i$. Then we have:
\begin{equation}\label{eq.1}
\begin{split}
S_{(\mathcal{R})}&=-\Tr\rho_{\mathcal{R}}\log\rho_{\mathcal{R}}\\
&=\Tr[\rho_{\mathcal{R}}(\sum_{i}s_in_i+\sum_{i\leq k,j\leq l,k,l}\lambda_{ijkl}c_i^\dag c_jc_k^\dag c_l)]+\log Z_{\mathcal{R}}\\
&=\sum_is_i\langle n_i\rangle+\sum_{i\leq k,j\leq l,k,l}\lambda_{ijkl}\langle c_i^\dag c_jc_k^\dag c_l\rangle+\log Z_{\mathcal{R}}.
\end{split}
\end{equation}
Then we can calculate these terms remaining up to the first order of $\lambda$.

We first consider the partition function:
\begin{equation}\label{eq.2}
\begin{split}
Z_{\mathcal{R}}&=\Tr e^{-S}\\
&=\Tr [\sum_m\frac{1}{m!}(-\sum_is_in_i-\sum_{i\leq k,j\leq l,k,l}\lambda_{ijkl}c_i^\dag c_jc_k^\dag c_l)^m].
\end{split}
\end{equation}
One can easily check that $\Tr(c_i^\dag c_jc_k^\dag c_l)$ will vanish unless it is diagonal, i.e., $i=j$ and $k=l$. Thus Eq.~\eqref{eq.2} can be rewritten as:
\begin{equation}\label{eq.3}
\begin{split}
Z_{\mathcal{R}}=&\Tr [\sum_m\frac{1}{m!}(-\sum_is_in_i-\sum_{i\leq j,j}\lambda_{iijj}n_in_j)^m]+\mathcal{O}(\lambda^2)\\
=&\Tr [\sum_m\frac{1}{m!}(-\sum_is_in_i)^m-\sum_m\frac{(-\sum_is_in_i)^{(m-1)}}{(m-1)!}\sum_{i\leq j,j}\lambda_{iijj}n_in_j]+\mathcal{O}(\lambda^2)\\
=&\Tr[e^{-\sum_is_in_i}(1-\sum_{i\leq j,j}\lambda_{iijj}n_in_j)]+\mathcal{O}(\lambda^2),
\end{split}
\end{equation}
where the first line is achieved since $[\sum_is_in_i,\sum_{i\leq j,j}\lambda_{iijj}n_in_j]=0$ with $[A,B]$ the commutator of operators $A$ and $B$. One can notice that the partition function consists of $Z_{{\mathcal{R}},0}=\Tr(e^{-\sum_is_in_i})$ and the trace of correction terms, i.e., $e^{-\sum_is_in_i}\sum_{ij}\lambda_{iijj}n_in_j$. For simplicity, we rewrite the partition function as:
\begin{equation}\label{eq.6}
\begin{split}
&\log{Z_{\mathcal{R}}}=\log{Z_{{\mathcal{R}},0}}+\sum_{i\leq j,j}\lambda_{iijj}f_{iijj}+\mathcal{O}(\lambda^2).
\end{split}
\end{equation}
Here we have defined $f_{iijj}=-\langle n_in_j\rangle_0$ with $\langle A\rangle_0=\Tr (e^{-S_0}A)$ the interaction-free expectation value of operator $A$. Meanwhile, one can easily check that the first term is:
\begin{equation}\label{eq.10}
\log{Z_{{\mathcal{R}},0}}=\sum_i\log(1+e^{-s_i})=-\log(1-M_{(\mathcal{R}),0}^{ii}),
\end{equation}
where $M_{(\mathcal{R}),0}$ represents the two-point correlation functions without interaction.

By now we have simplified the first term in Eq.~\eqref{eq.1}. The next to derive are the four-point correlation functions, i.e., the second term in Eq.~\eqref{eq.1}:
\begin{equation}\label{eq.7}
\begin{split}
&\lambda_{ijkl}\langle c_i^\dag c_jc_k^\dag c_l\rangle=\lambda_{ijkl}\langle c_i^\dag c_jc_k^\dag c_l\rangle_0+\mathcal{O}(\lambda^2).
\end{split}
\end{equation}
For the same reason as in Eq.~\eqref{eq.2}, we can see the surviving terms of the second terms in eq.~\eqref{eq.1} should be $\sum_{i\leq j,j}\lambda_{iijj}\langle n_in_j\rangle$.
As stated in Ref.~\cite{pachos2022quantifying}:
\begin{equation}\label{eq.8}
\langle n_in_j\rangle=-\frac{\partial \log{Z_{\mathcal{R}}}}{\partial \lambda_{iijj}}.
\end{equation}
So with the expression of partition function in Eq.~\eqref{eq.6}, we have:
\begin{equation}\label{eq.9}
\begin{split}
\sum_{i\leq j,j}\lambda_{iijj}\langle n_in_j\rangle&=-\sum_{i\leq j,j}\lambda_{iijj}f_{iijj}=\log{Z_{{\mathcal{R}},0}}-\log{Z_{\mathcal{R}}}+\mathcal{O}(\lambda^2).
\end{split}
\end{equation}

The last part remaining are the quadratic terms in Eq.~\eqref{eq.1}. Since $M_{(\mathcal{R}),0}^{ii}=1/(1+e^{s_i})$, the quadratic terms in Eq.~\eqref{eq.1} can be rewritten as:
\begin{equation}\label{eq.11}
\begin{split}
&\sum_is_i\langle n_i\rangle=\sum_i\log[\frac{1-M_{(\mathcal{R}),0}^{ii}}{M_{(\mathcal{R}),0}^{ii}}]M_{(\mathcal{R})}^{ii}\\
&=\sum_i[M_{(\mathcal{R})}^{ii}\log(1-M_{(\mathcal{R}),0}^{ii})-M_{(\mathcal{R})}^{ii}\log{M_{(\mathcal{R}),0}^{ii}}].
\end{split}
\end{equation}

Taking all these derivations together, we can rewrite Eq.~\eqref{eq.1} as:
\begin{equation}\label{eq.12}
\begin{split}
S_{(\mathcal{R})}=&\sum_is_i\langle n_i\rangle+\sum_{i\leq k,j\leq l,k,l}\lambda_{ijkl}\langle c_i^\dag c_jc_k^\dag c_l\rangle+\log Z_{\mathcal{R}}\\
=&\sum_i[M_{(\mathcal{R})}^{ii}\log(1-M_{(\mathcal{R}),0}^{ii})-M_{(\mathcal{R})}^{ii}\log{M_{(\mathcal{R}),0}^{ii}}]-\log{Z_{\mathcal{R}}}+\log{Z_{{\mathcal{R}},0}}+\log{Z_{\mathcal{R}}}+\mathcal{O}(\lambda^2)\\
=&\sum_i[M_{(\mathcal{R})}^{ii}\log(1-M_{(\mathcal{R}),0}^{ii})-M_{(\mathcal{R})}^{ii}\log{M_{(\mathcal{R}),0}^{ii}}-\log(1-M_{(\mathcal{R}),0}^{ii})]+\mathcal{O}(\lambda^2)\\
=&-\sum_i[(1-M_{(\mathcal{R})}^{ii})\log(1-M_{(\mathcal{R}),0}^{ii})+M_{(\mathcal{R})}^{ii}\log{M_{(\mathcal{R}),0}^{ii}}]+\mathcal{O}(\lambda^2).
\end{split}
\end{equation}
What's more, up to the first order of $\lambda$, the correlation function should always satisfy:
\begin{equation}\label{eq.13}
M_{(\mathcal{R})}^{ii}=M_{(\mathcal{R}),0}^{ii}+M_{(\mathcal{R}),1}^{ii}+\mathcal{O}(\lambda^2)=M_{(\mathcal{R}),0}^{ii}+M_{(\mathcal{R})}^{'ii},
\end{equation}
where the first order correction $M_{(\mathcal{R}),1}^{ii}$ is linear to $\lambda$ and $M_{(\mathcal{R})}^{'ii}:=M_{(\mathcal{R})}^{ii}-M_{(\mathcal{R}),0}^{ii}$. Then
\begin{equation}\label{eq.14}
\begin{split}
S_{(\mathcal{R})}=&-\sum_i[(1-M_{(\mathcal{R})}^{ii})\log(1-M_{(\mathcal{R}),0}^{ii})+M_{(\mathcal{R})}^{ii}\log{M_{(\mathcal{R}),0}^{ii}}]+\mathcal{O}(\lambda^2)\\
=&-\sum_i[(1-M_{(\mathcal{R})}^{ii})\log(1-M_{(\mathcal{R})}^{ii}+M_{(\mathcal{R})}^{'ii})+M_{(\mathcal{R})}^{ii}\log(M_{(\mathcal{R})}^{ii}-M_{(\mathcal{R})}^{'ii})]+\mathcal{O}(\lambda^2)\\
=&-\sum_i[(1-M_{(\mathcal{R})}^{ii})\log(1-M_{(\mathcal{R})}^{ii})+M_{(\mathcal{R})}^{ii}\log{M_{(\mathcal{R})}^{ii}}+(1-M_{(\mathcal{R})}^{ii})\frac{M_{(\mathcal{R})}^{'ii}}{1-M_{(\mathcal{R})}^{ii}}-M_{(\mathcal{R})}^{ii}\frac{M_{(\mathcal{R})}^{'ii}}{M_{(\mathcal{R})}^{ii}}]+\mathcal{O}(\lambda^2)\\
=&-\sum_i[(1-M_{(\mathcal{R})}^{ii})\log(1-M_{(\mathcal{R})}^{ii})+M_{(\mathcal{R})}^{ii}\log{M_{(\mathcal{R})}^{ii}}]+\mathcal{O}(\lambda^2)\\
=&-\Tr[(1-M_{(\mathcal{R})})\log(1-M_{(\mathcal{R})})+M_{(\mathcal{R})}\log{M_{(\mathcal{R})}}]+\mathcal{O}(\lambda^2).
\end{split}
\end{equation}

Now we have established the relationship between VNE and two-point correlation functions. However, remember that our starting point is that the reduced density matrix is an exponential, i.e., $\rho_{(\mathcal{R})}=e^{-S_0-S_I}$, which is usually not the case.
Generally, a reduced density matrix is not in an exponential form so that $\log\rho_{\mathcal{R}}$ is hard to simplify. Fortunately, nevertheless, based on the fact that the reduced density matrix of a fermion Gaussian state is still a fermion Gaussian state, we can generally expand the reduced density matrix of a weakly interacting system around fermion Gaussian state:
\begin{equation}\label{eq.16}
\rho_{\mathcal{R}}=\frac{e^{-\sum_is'_in_i}(1+f^{(l)}({\Lambda},\{c^\dag,c\}_{(c)}))+\mathcal{O}(\Lambda^{l+1})}{Z_{\mathcal{R}}}.
\end{equation}
Here $f$ is a function of fermion operators and interaction strength, the superscript $l$ denotes the leading order and the subscript $(c)$ denotes that the system is charge-conserved. $f$ is proportional to the leading order of interaction strength and vanishes without interaction. For example, as we show in the main text, for a subsystem of a thermal equilibrium system, $f=(\sum_{ijkl}\Lambda_{ijkl}g_{ijkl}^{(c)})^l$.
Then with Eq.~\eqref{eq.16}, VNE becomes:
\begin{equation}\label{eq.17}
\begin{split}
S_{(\mathcal{R})}=&-\Tr{\rho_{\mathcal{R}}\log{\rho_{\mathcal{R}}}}\\
=&\Tr{\rho_{\mathcal{R}}[\sum_is'_in_i-\log{(1+f)}+\mathcal{O}(\lambda^{l+1})]}+\log{Z_{\mathcal{R}}}\\
=&\Tr{\rho_{\mathcal{R}}[\sum_is'_in_i-f+\mathcal{O}(\lambda^{l+1})]}+\log{Z_{\mathcal{R}}}\\
=&\sum_is'_i\langle n_i\rangle-\langle f\rangle+\log{Z_{\mathcal{R}}}+\mathcal{O}(\lambda^{l+1})\\
=&\sum_is'_i\langle n_i\rangle-\langle f_{diag}\rangle_0+\log{Z_{\mathcal{R}}}+\mathcal{O}(\lambda^{l+1}).
\end{split}
\end{equation}
The subscript ``diag'' in the last line denotes the diagonal terms and the non-diagonal terms are dropped. The subscript ``$0$'' in $\langle f_{diag}\rangle_0$ comes from the fact that $f$ function itself is the leading order correction so we drop away the higher order correlation functions.

With similar steps as in the previous proof, the partition function is:
\begin{equation}\label{eq.18}
\begin{split}
Z_{\mathcal{R}}=&\Tr[{e^{-\sum_is'_in_i}(1+f_{diag})+\mathcal{O}(\lambda^{l+1})}]\\
=&Z_{{\mathcal{R}},0}+\Tr[e^{-\sum_is'_in_i}f_{diag}]+\mathcal{O}(\lambda^{l+1})\\
=&Z_{{\mathcal{R}},0}+Z_{{\mathcal{R}},0}\langle f_{diag}\rangle_0+\mathcal{O}(\lambda^{l+1}).
\end{split}
\end{equation}
Therefore,
\begin{equation}\label{eq.18}
\begin{split}
\log{Z_{\mathcal{R}}}=&\log{Z_{{\mathcal{R}},0}}+\langle f_{diag}\rangle_0+\mathcal{O}(\lambda^{l+1}).
\end{split}
\end{equation}

So we have achieved the VNE with the same structure as in Eq.~\eqref{eq.12}. As a result, the final result is the same as before:
\begin{equation}\label{res}
S_{(\mathcal{R})}=-\Tr[(1-M_{(\mathcal{R})})\log(1-M_{(\mathcal{R})})+M_{(\mathcal{R})}\log M_{(\mathcal{R})}]+\mathcal{O}(\lambda^{l+1})
\end{equation}

To conclude, while two-point correlation function cannot describe the interacting system due to the failure of Wick's theorem, we can see that, however, the von Neumann entropy is determined only by the two-point correlation functions up to leading order of weak interactions. What's more, Eq.~\eqref{eq.14} and Eq.~\eqref{res} do not contain details of interactions that are hard to detect, which allows us to further propose a general proposal to measure VNE production.

\subsection{Beyond perturbation theory: Saddle-point Approximation}\label{III}
In the previous section, we treat the interaction as a perturbation, which is, however, not always the case in real systems. To go beyond perturbation theory, we turn to mean field theory. If mean field theory can be applied in a system, we can expand the density matrix of this system around its saddle point. So in this section, we will handle the interaction with Hubbard-Stratonovich (H-S) transformation.

By applying H-S transformation, the interaction term of fermions can be written in a quadratic form and coupled to a bosonic field $\varphi$:
\begin{equation}
\begin{split}
e^{\tilde{c}_i^\dag \tilde{c}_j\tilde{c}_k^\dag \tilde{c}_l}= \int D[{\varPhi}] e^{\tilde{c}_i^\dag \tilde{c}_j \tilde{c}_k^\dag \tilde{c}_l-\varPhi_{ij}\varPhi_{kl}}/Z_{\varPhi},\\
\varPhi_{ij}\rightarrow \varphi_{ij}-\tilde{c}_i^\dag \tilde{c}_j,\\
\downarrow\\
e^{\tilde{c}_i^\dag \tilde{c}_j \tilde{c}_k^\dag \tilde{c}_l}= \int D[{\varphi} ]e^{\varphi_{ij}\tilde{c}_k^\dag \tilde{c}_l+\varphi_{kl}\tilde{c}_i^\dag \tilde{c}_l-\varphi_{ij}\varphi_{kl}}/Z_{\varphi}.
\end{split}
\end{equation}
After the H-S transformation, the density matrix of a thermal equilibrium system reads (we will discuss the non-equilibrium case in Sec.~\ref{V}):
\begin{equation}\label{HSrho1}
\begin{split}
\rho&=e^{-\beta(\sum_{ij}\epsilon_{ij}\tilde{c}_i^\dag \tilde{c}_j-\sum_{ijkl}\lambda_{ijkl}\tilde{c}_i^\dag \tilde{c}_j\tilde{c}_k^\dag \tilde{c}_l)}/Z\\
&=\int D[{\varphi}]e^{-\beta[\sum_{ij}\epsilon_{ij}\tilde{c}_i^\dag \tilde{c}_j-\sum_{ijkl}\lambda_{ijkl}(\varphi_{ij}\tilde{c}_k^\dag \tilde{c}_l+\varphi_{kl}\tilde{c}_i^\dag \tilde{c}_j-\varphi_{ij}\varphi_{kl})]}/(ZZ_{\varphi}).
\end{split}
\end{equation}
Here and in the following of this section, we will hide the notation ``$i\leq k,\ j\leq l$" for simplicity.
Following a standard method, one can find out the saddle point $\varphi_0$ by solving $\partial_{\varphi} Z_{eff}|_{\varphi_0}=0$ where $Z_{eff}$ is defined as:
\begin{equation}\label{zeff}
Z_{eff}(\varphi)=\Tr_{\tilde{c}}e^{-\beta[\sum_{ij}\epsilon_{ij}\tilde{c}_i^\dag \tilde{c}_j-\sum_{ijkl}\lambda_{ijkl}(\varphi_{ij}\tilde{c}_k^\dag \tilde{c}_l+\varphi_{kl}\tilde{c}_i^\dag \tilde{c}_j-\varphi_{ij}\varphi_{kl})]}.
\end{equation}
Here $\Tr_{\tilde{c}}$ represents the tracing out of fermionic degrees of freedom.
With the saddle-point solution, the density matrix \eqref{HSrho1} reads:
\begin{equation}\label{HSrho}
\begin{split}
\rho&=\int D[{\varphi}]e^{-\beta[\sum_{ij}\epsilon_{ij}\tilde{c}_i^\dag \tilde{c}_j-\sum_{ijkl}\lambda_{ijkl}(\varphi_{ij}\tilde{c}_k^\dag \tilde{c}_l+\varphi_{kl}\tilde{c}_i^\dag \tilde{c}_j-\varphi_{ij}\varphi_{kl})]}/(ZZ_{\varphi})\\
&=\int D[{\phi}]e^{-\beta[\sum_{ij}\epsilon'_{ij}\tilde{c}_i^\dag \tilde{c}_j-\sum_{ijkl}\lambda_{ijkl}(\phi_{ij}\tilde{c}_k^\dag \tilde{c}_l+\phi_{kl}\tilde{c}_i^\dag \tilde{c}_j-\phi_{ij}\phi_{kl})]}/(ZZ_{\varphi}),
\end{split}
\end{equation}
where $\epsilon'_{ij}=\epsilon_{ij}-\sum_{kl}(\lambda_{ijkl}+\lambda_{klij})\varphi_{0,ij}$, and $\phi_{ij}=\varphi_{ij}-\varphi_{0,ij}$. $\phi$ represents the bosonic field fluctuations around the saddle point. Here we have absorbed the saddle-point contribution into $\epsilon'$ and, at the saddle point $\varphi=\varphi_0$, the density matrix is exactly non-interacting .

Note that we are interested in the von Neumann entropy of the fermion part, so we should integrate out the bosonic field in Eq.~\eqref{HSrho} first. In the standard H-S transformation, one can approximately keep the Gaussian fluctuation of the bosonic field around a saddle point. In this approximation, the premises are that the saddle point is stable and the bosonic fluctuation is weak around the saddle point. As a consequence, in Eq.~\eqref{HSrho}, we can expand $\rho$ around $\phi=0$ and keep only the Gaussian terms, which results to quartic interaction terms:
\begin{equation}\label{rhof}
\begin{split}
\rho&=\int D[{\phi}]e^{-\beta[\sum_{ij}\epsilon'_{ij}\tilde{c}_i^\dag \tilde{c}_j-\sum_{ijkl}\lambda_{ijkl}(\phi_{ij}\tilde{c}_k^\dag \tilde{c}_l+\phi_{kl}\tilde{c}_i^\dag \tilde{c}_j-\phi_{ij}\phi_{kl})]}/(ZZ_{\varphi})\\
&=\int D[{\phi}]e^{-\beta[\sum_{ij}\epsilon'_{ij}\tilde{c}_i^\dag \tilde{c}_j-\sum_{ijkl}(-\lambda_{ijkl}\phi_{ij}\phi_{kl})]}\{1+\sum_{ijkl}\phi_{ij}\phi_{kl}G_{ijkl}\}/(ZZ_{\varphi})\\
&=e^{-\beta[\sum_{ij}\epsilon'_{ij}\tilde{c}_i^\dag \tilde{c}_j]}(1+\sum_{ijkl}\tilde{G}_{ijkl})/\tilde{Z},
\end{split}
\end{equation}
where $G_{ijkl}$s and $\tilde{G}_{ijkl}$s are fermionic terms and they are related through tracing out of the bosonic field:
\begin{equation}
\tilde{G}_{ijkl}=\int D[\phi] e^{-\beta\sum_{ijkl}\lambda_{ijkl}\phi_{ij}\phi_{kl}}\phi_{ij}\phi_{kl}G_{ijkl}/Z_\varphi.
\end{equation}
In the second line of \eqref{rhof}, $\phi\phi G$s come from the expansion around $\phi=0$ up to the Gaussian order (the linear terms vanish due to the saddle-point condition). Here we do not show the explicit expression of $G_{ijkl}$ and $\tilde{G}_{ijkl}$ since with the non-commutative properties between fermionic terms, $G_{ijkl}$ is usually complicated. Actually, according to Zassenhaus formula, it is easy to check that $G_{ijkl}$ usually contains quadratic and quartic terms, which is, however, hard to be written in an explicit expression. Nevertheless, we will show later that we don't need the expression of $\tilde{G}_{ijkl}$ and just keep in mind that it comes from the Gaussian fluctuations of bosonic field.

Then we can trace out the Hilbert space of the region $\bar{\mathcal{R}}$, $\mathcal{H}_{\bar{\mathcal{R}}}$, to obtain the reduced density matrix of $\mathcal{R}$, i.e., $\rho_\mathcal{R}=\Tr_{\mathcal{H}_{\bar{\mathcal{R}}}} \rho$. Since the non-interacting part of Eq.~\eqref{rhof} turns out to be still quadratic after tracing out a subspace of the whole Hibert space, we have:
\begin{equation}\label{reduced}
\begin{split}
\rho_\mathcal{R}&=\Tr_{\mathcal{H}_{\bar{\mathcal{R}}}}\rho\\
&=e^{-\beta(\sum_{i,j\in\mathcal{H}_\mathcal{R}}\epsilon''_{ij}\tilde{c}_i^\dag \tilde{c}_j)}\{1+\sum_{ijkl}\tilde{G}'_{ijkl}\}/Z_\mathcal{R}\\
&=e^{-\beta(\sum_{i\in\mathcal{H}_\mathcal{R}}\tilde{\epsilon}_{i}c_i^\dag c_i)}\{1+\sum_{ijkl}G'_{ijkl}\}/Z_\mathcal{R}.
\end{split}
\end{equation}
Here the second line comes from the tracing out of the Hilbert space of region $\bar{\mathcal{R}}$, i.e., $\Tr_{\mathcal{H}_{\bar{\mathcal{R}}}}e^{-\beta\sum_{ij}\epsilon'_{ij}\tilde{c}_i^\dag \tilde{c}_j}\tilde{G}_{ijkl}/Z_\mathcal{R}=e^{-\beta\sum_{i,j\in\mathcal{H}_\mathcal{R}}\epsilon''_{ij} \tilde{c}_i^\dag \tilde{c}_j}\tilde{G}'_{ijkl}/Z_\mathcal{R}$. We have diagonalized the terms $\sum_{ij}\epsilon'_{ij}\tilde{c}_i^\dag \tilde{c}_j$ in the second line of \eqref{reduced} to get the last one and label the new fermion operators $c$ and $c^\dag$.
The terms $G'$ are the leading order of the correction of the reduced density matrix. Apparently, this reduced density matrix is similar to Eq.~\eqref{eq.16}. Therefore, the subsequent steps are the same as in Sec.~\ref{II} and we don't need the expression of $G'_{ijkl}$. As a consequence, we prove that the follwing relationship still hold around the saddle point up to the leading order of quantum fluctuation:
\begin{equation}\label{vnesaddle}
S_{(\mathcal{R})}=-\Tr[(1-M_{(\mathcal{R})})\log(1-M_{(\mathcal{R})})+M_{(\mathcal{R})}\log M_{(\mathcal{R})}]
\end{equation}

\subsection{Entanglement measurement through transport with interaction}\label{IV}
We have proven that transport can tell us the spectrum of two-point correlation functions in non-interacting multi-partite systems in Sec.~\ref{I}. However, our starting point, Eq.~\eqref{generate}, has not been proven to be valid in interacting systems. To deal with a interaction system, we have to restart with the definition of the generating function in full counting statistics:
\begin{equation}\label{genedef}
\chi(\lambda)=\Tr[\rho_\mathcal{R} e^{i\lambda(\sum_{t\in \mathcal{H}_\mathcal{R}}c_t^\dag c_t-N_\mathcal{R}^0)}]
\end{equation}
Obviously, Eq.~\eqref{genedef} is exactly the same as Levitov's definition, Eq.~\eqref{generating}. We will discuss the perturbation case as an example and use reduced density matrix in the form similar to Eq.~\eqref{eq.16}:
\begin{equation}\label{generho}
\rho_\mathcal{R}=\frac{e^{-\sum_{t\in \mathcal{H}_\mathcal{R}}s_tc_t^\dag c_t}(1-f)}{Z_\mathcal{R}}.
\end{equation}
Here $f$ is the leading order correction of the reduced density matrix from the weak interaction and from now on, we just omit the higher order correction $\mathcal{O}(\Lambda^{l+1})$ for simplicity. The weak-fluctuation case will be discussed later. By substituting Eq.~\eqref{generho} into Eq.~\eqref{genedef}, we have:
\begin{equation}\label{chiint}
\begin{split}
\log\chi(\lambda)&=-i\lambda N_\mathcal{R}^0+\log\frac{\Tr[e^{-\sum_{t\in\mathcal{H}_\mathcal{R}}(s_t-i\lambda)c_t^\dag c_t}(1-f)]}{\Tr[e^{-\sum_{t\in\mathcal{H}_\mathcal{R}}s_tc_t^\dag c_t}(1-f)]}\\
&=-i\lambda N_\mathcal{R}^0+\log\frac{\Tr[e^{-\sum_{t\in\mathcal{H}_\mathcal{R}}(s_t-i\lambda)c_t^\dag c_t}(1-f)]}{\Tr[e^{-\sum_{t\in\mathcal{H}_\mathcal{R}}(s_t-i\lambda)c_t^\dag c_t}]}-\log\frac{\Tr[e^{-\sum_{t\in\mathcal{H}_\mathcal{R}}s_tc_t^\dag c_t}(1-f)]}{\Tr[e^{-\sum_{t\in\mathcal{H}_\mathcal{R}}s_tc_t^\dag c_t}]}+\log\frac{\Tr[e^{-\sum_{t\in\mathcal{H}_\mathcal{R}}(s_t-i\lambda)c_t^\dag c_t}]}{\Tr[e^{-\sum_{t\in\mathcal{H}_\mathcal{R}}s_tc_t^\dag c_t}]}\\
&=\log\chi_0(\lambda)+\log(1-\langle f\rangle_\lambda)-\log(1-\langle f\rangle_0)\\
&=\log\chi_0(\lambda)-\langle f\rangle_\lambda+\langle f\rangle_0,
\end{split}
\end{equation}
where $\log\chi_0(\lambda)=\log\Tr [e^{-\sum_{t\in\mathcal{H}_\mathcal{R}}(s_t-i\lambda)c_t^\dag c_t}]-\log{\Tr [e^{-\sum_{t\in\mathcal{H}_\mathcal{R}}s_tc_t^\dag c_t}}]-i\lambda N_\mathcal{R}^0$ is the generating function without interaction. In addition, for simplicity, we have defined $\langle f\rangle_\lambda:=[\Tr e^{-\sum_{t\in\mathcal{H}_\mathcal{R}}(s_t-i\lambda)c_t^\dag c_t}f]/[{\Tr e^{-\sum_{t\in\mathcal{H}_\mathcal{R}}(s_t-i\lambda)c_t^\dag c_t}}]$ and $\langle f\rangle_0:=[\Tr e^{-\sum_{t\in\mathcal{H}_\mathcal{R}}s_tc_t^\dag c_t}f]/[{\Tr e^{-\sum_{t\in\mathcal{H}_\mathcal{R}}s_tc_t^\dag c_t}}]$. Since we have diagonalized the quadratic part in Eq.~\eqref{generho}, the expectation values of $f$, both $\langle f\rangle_0$ and $\langle f\rangle_\lambda$ will vanish except for the diagonal terms, $f_{diag}$, as we have argued in Eq.~\eqref{eq.3}. Therefore, in the following derivation, we will only keep the diagonal terms, $f_{diag}$. Remarkably, one should not worry that if $\langle A\rangle_\lambda$ is well-defined: In Eq.~\eqref{integ}, we have limited $0\leq z\leq 1$ which means $i\lambda$ is real and, as a consequence, we can treat $\sum_ti\lambda n_t$ as an effective action and $\langle A\rangle_\lambda$ as an effective expectation value.

Physically, the interaction terms are usually quartic, so if the system is a thermal state, we first assume that the $f$ terms satisfy (the non-equilibrium case will be discussed in Sec.~\ref{V}):
\begin{equation}\label{quartic}
f_{diag}=\sum_{k<l,l}\Gamma_{kl}c_k^\dag c_kc_l^\dag c_l.
\end{equation}
Here we only keep quartic terms which satisfy $k\neq l$ since $c_k^\dag c_kc_k^\dag c_k=c_k^\dag c_k$ is no longer quartic. Eq.~\eqref{quartic} is valid when there are interactions in the bulk since there will always be quartic terms in the reduced density matrix after tracing out the Hilbert space of region $\bar{\mathcal{R}}$. With these assumptions, Eq.~\eqref{chiint} becomes:
\begin{equation}\label{geneint}
\begin{split}
\log\chi(\lambda)=\log\chi_0(\lambda)-\sum_{k<l,l}\Gamma_{kl}\langle c_k^\dag c_kc_l^\dag c_l\rangle_\lambda+\sum_{k<l,l}\Gamma_{kl}\langle c_k^\dag c_kc_l^\dag c_l\rangle_0.
\end{split}
\end{equation}
Eq.~\eqref{geneint} can be simplified to be:
\begin{equation}\label{chibulk}
\log\chi(\lambda)=\sum_i\log(z-M_{(\mathcal{R})}^{ii})-\sum_{k<l,l}\frac{\Gamma_{kl}M_{(\mathcal{R}),0}^{kk}M_{(\mathcal{R}),0}^{ll}(1-M_{(\mathcal{R}),0}^{kk})(1-M_{(\mathcal{R}),0}^{ll})}{(z-M_{(\mathcal{R}),0}^{kk})(z-M_{(\mathcal{R}),0}^{ll})}+Irr,
\end{equation}
where $z=(1-e^{i\lambda})^{-1}$ and $Irr=N_{\mathcal{R}}^0\log [z/(z-1)]-\sum_t\log[(1-M_{(\mathcal{R}),0}^{tt})z]$ being the irrelevant terms when handling VNE~\cite{klich2009quantum}. In the following, we will drop these irrelevant terms. In Eq.~\eqref{chibulk} we have used the fact that $M_{(\mathcal{R}),0}^{ii}=(e^{s_i}+1)^{-1}$ and 
$\log\chi_0=\sum_i\log(z-M_{(\mathcal{R}),0}^{ii})$. What's more, during the derivation, we have used the following property of two-point correlation function:
\begin{equation}
\begin{split}
M_{(\mathcal{R})}^{ii}&=\frac{\Tr[e^{-\sum_ts_tc_t^\dag c_t} (1-f_{diag})c_i^\dag c_i]}{\Tr[e^{-\sum_ts_tc_t^\dag c_t} (1-f_{diag})]}\\
&=\langle c_i^\dag c_i\rangle_0-\sum_{k<l,l}\Gamma_{kl}\langle c_k^\dag c_k c_l^\dag c_lc_i^\dag c_i\rangle_0+\sum_{k<l,l}\Gamma_{kl}\langle c_k^\dag c_kc_l^\dag c_l\rangle_0\langle c_i^\dag c_i\rangle_0\\
&=M_{(\mathcal{R}),0}^{ii}-\sum_{k<i}\Gamma_{ki}M_{(\mathcal{R}),0}^{kk}M_{(\mathcal{R}),0}^{ii}(1-M_{(\mathcal{R}),0}^{ii})-\sum_{l>i}\Gamma_{il}M_{(\mathcal{R}),0}^{ll}M_{(\mathcal{R}),0}^{ii}(1-M_{(\mathcal{R}),0}^{ii}).
\end{split}
\end{equation}

We can see that in Eq.~\eqref{chibulk}, the first term can give us the spectrum of two-point correlation function through $\Im\partial_z\log\chi$. However, the second term is the unwanted deviation since it depends on the details of interaction which is hard to detect. To make this term infinitely small, we have to approach the limit where $M_{(\mathcal{R}),0}^{ii}\rightarrow 0$ or $M_{(\mathcal{R}),0}^{ii}\rightarrow 1$ for all $i\in \mathcal{H}_{\mathcal{R}}$. Since we have diagonalized the reduced density matrix, $k$ state is the diagonal basis and at zero temperature, $M_i=\theta(\beta E_F-s_i)$ with $\theta(x)$ the step function. Therefore, if the final $\rho_\mathcal{R}$ is in thermal equilibrium, we have to limit the system at sufficiently low temperature, $T(t_f)\to 0$. The non-equilibrium case will be discussed in Sec.~\ref{V}.

In addition to the bulk interaction, there is a better case: boundary-interaction case. Unlike the above bulk interaction, interactions can be introduced by a quantum dot at the boundary. For example, with a quantum dot deep in Coulomb blockage valley, the effective weak interaction term reads $H_I=\sum_{i\in \mathcal{H}_{lead1}}\sum_{j\in \mathcal{H}_{lead2}}\sum_{k,l\in\mathcal{H}_{dot}} (J_{ijkl}c_i^\dag c_jd_k^\dag d_l+J_{jikl}c_ic_j^\dag d_k^\dag d_l)$ with $d$ and $d^\dag$ the annihilation and creation operators of the excitations in dot. If the system is in a thermal equilibrium state (we will discuss the non-equilibrium case in Sec.~\ref{V}), we have $\rho=e^{-\beta(H_0+H_I)}$ with $H_0$ the quadratic part without coupling between the two leads. To obtain a reduced density matrix of lead 1, we need to expand $\rho$ up to the leading order of interaction and trace out the Hilbert space of lead 2 and dot. Since $\Tr_{\mathcal{H}_{lead2}}c_j=0$, the first order correction vanishes. The second order expansion, in contrast, leads to a quadratic $f$ as the leading order correction. To show this, as an example, suppose we have the density matrix of a whole system in 3-terminal topological Kondo effect:
\begin{equation}
\begin{split}
\rho&=e^{-\beta(H_0+\sum_{t1,t2=\{1,2,3\}}J_{t_1t_2}\gamma_{t_1}\gamma_{t_2}c_{t_1}^\dag c_{t_2})}/Z\\
&:=e^{-\beta(H_0+H_T)},
\end{split}
\end{equation}
where $\gamma_{t_i}$ is the Majorana operator coupled with the $t_i$-th lead and $c_{t_i}$ is the annihilation operator at the end of the $t_i$-th lead. To get the reduced density matrix of lead 1, we can expand the density matrix as:
\begin{equation}\label{trace}
\begin{split}
\rho_1&=\Tr_{\mathcal{H}_\gamma}\Tr_{\mathcal{H}_2}\Tr_{\mathcal{H}_3}\rho\\
&=\Tr_{\mathcal{H}_\gamma}\Tr_{\mathcal{H}_2}\Tr_{\mathcal{H}_3}[e^{-\beta H_0}e^{-\beta H_T}\prod_xe^{\beta H_{C,x}}]/Z\\
&=\Tr_{\mathcal{H}_\gamma}\Tr_{\mathcal{H}_2}\Tr_{\mathcal{H}_3}[e^{-\beta H_0}(1-\beta H_T-\sum_x\beta H_{C,x}+\frac{\beta^2}{2}H_T^2+\frac{\beta^2}{2}(\sum_xH_{C,x})^2)],
\end{split}
\end{equation}
where $H_{C,x}$ is linear to $J_{t_it_j}$ and comes from the commutative relation between $H_0$ and $H_T$, which can be obtained from Zassenhaus formula: $H_{C,x}\sim [H_0,[H_0,\dots[H_0,H_T]\dots]]+\dots$ and there are $x$ commutators in $H_{C,x}$. Since $[c_i^\dag c_j, c_k^\dag c_l]=c_i^\dag c_l\delta_{jk}-c_k^\dag c_j\delta_{il}$ for fermionic operators, one can easily check that $H_C$ is quartic and contains only one creation/annihilation operator (CAO) of each lead. So in Eq.~\eqref{trace}, $H_T$, $H_C$ vanish and $H_T^2$, $H_C^2$ survive. Since $H_T^2$, $H_C^2$ contains two CAO for each lead, after tracing out the Hilbert space of other leads and dot, the remaining correction for $\rho_1$ is quadratic. As a result, with the interaction at the boundary, the correction term in Eq.~\eqref{generho} is now:
\begin{equation}\label{impurityint}
f_{diag}=\sum_k\Gamma'_kc_k^\dag c_k.
\end{equation}
We can substitute Eq.~\eqref{impurityint} into Eq.~\eqref{genedef}. Note that now $M_{(\mathcal{R})}^{ii}=M_{(\mathcal{R}),0}^{ii}+\Gamma'_i(M_{(\mathcal{R}),0}^{ii})^2-\Gamma'_iM_{(\mathcal{R}),0}^{ii}$, so we get:
\begin{equation}\label{chiedge}
\begin{split}
\log\chi(\lambda)&=\log \chi_0(\lambda)-\sum_k\Gamma'_k\langle c_k^\dag c_k\rangle_\lambda+\sum_k\Gamma'_k\langle c_k^\dag c_k\rangle_0\\
&=\sum_i\log[z-M_{(\mathcal{R}),0}^{ii}-\Gamma'_i(z-1)M_{(\mathcal{R}),0}^{ii}+\Gamma'_iM_{(\mathcal{R}),0}^{ii}(z-M_{(\mathcal{R}),0}^{ii})]\\
&=\sum_i\log(z-M_{(\mathcal{R})}^{ii}).
\end{split}
\end{equation}
The derivation method is similar to that in the bulk-interaction case. As a result, the spectrum of two-point correlation function is achieved through $\mu(z)=\Im\partial_z\log\chi$ just as in the non-interacting case, which protects the validity of Eq.~\eqref{result} in the main text.

The proof above can be similarly applied to the saddle-point approximation case. Note that Eq.~\eqref{generho} and Eq.~\eqref{chiint} are also valid in this case. So the only thing we need to do is to analyze the correction term $f$. Apparently, with interaction in the bulk, $f$ is generally quartic. If the interaction is located at the boundary, $\tilde{G}$ term in Eq.~\eqref{rhof} comes from Gaussian fluctuation which is similar to $H_C^2$ and $H_T^2$ in Eq.~\eqref{trace}. Following a similar argument, we can easily prove that $f$ is quadratic with interaction at the boundary. As a consequence, the results in weakly interacting case, Eqs.~\eqref{chibulk} and~\eqref{chiedge}, also hold in weakly fluctuating case.

\subsection{Beyond thermal state}\label{V}
In this section, we will prove that the above results, Eqs.~\eqref{res},~\eqref{vnesaddle},~\eqref{chibulk} and~\eqref{chiedge} are also valid with a non-equilibrium state. We concentrate on a system whose initial state is thermal equilibrium and the density matrix is $\rho_0=e^{-S_0-S_I}/Z$ where $S_0$ is quadratic and $S_I$ is quartic. In this density matrix, $S_I$ arises from the interaction, i.e., $S_I\propto\beta H_I$. Below, we assume that at the initial time $t_i=0$, the evolution begins and the Hamiltonian of this evolving system is $H=H_2+H_4$ with $H_2$ the quadratic part and $H_4$ the quartic part. Then the density matrix of the whole system becomes:
\begin{equation}\label{neqrho}
\rho=e^{-iHt}\rho_0 e^{iHt}.
\end{equation}
We will derive the expression of reduced density matrix with Eq.~\eqref{neqrho} in weakly interacting and fluctuating case, respectively. The key target in this section is to prove that
the reduced density matrix can be written in the form of:
\begin{equation}\label{rhogeneral}
\rho_{\mathcal{R}}=e^{-\sum_is_in_i}(1+f)/Z,
\end{equation}
just as in Eq.~\eqref{eq.16}. Here we call $f$ the correction term. As long as we achieve Eq.~\eqref{rhogeneral}, Eq.~\eqref{theroem_1} in the main text can be achieved as a result (see Sec.~\ref{II}). In addition, to prove the validity of entanglement measurement through transport, we need to analyze the form of $f$ in the presence of interaction at the boundary or in the bulk, as we did in Sec.~\ref{IV}.

\subsubsection{Weakly Interacting Case}\label{VA}
By defining an unitary evolving operator $U=e^{i(H_2+H_4)t}$, we can rewrite the density matrix as:
\begin{equation}\label{noneqwp}
\begin{split}
\rho&=U^\dag\rho_0 U=\frac{U^\dag e^{-S_0-S_I}U}{Z}\\
&=\sum_m\frac{U^\dag (-S_0-S_I)^mU}{m!Z}\\
&=\frac{e^{U^\dag(-S_0-S_I)U}}{Z},
\end{split}
\end{equation}
where we get the last line by inserting $(m-1)$ copies of $UU^\dag$ inside the term $(-S_0-S_I)^m$. If the contribution of action $iH_4t$ is sufficiently small and the perturbation approximation works, we can expand the evolving operator up to the first order:
\begin{equation}\label{evovling}
\begin{split}
U^\dag&=e^{-i(H_2+H_4)t}=e^{-iH_2t}e^{-iH_4t}e^{-[H_2,H_4]t^2}\cdots\\
&=e^{-iH_2t}(1-i(H_4-i[H_2,H_4]t+\cdots)t)\\
&=e^{-iH_2t}(1-i\tilde{H}_4t)\\
&=e^{-iH_2t}e^{-i\tilde{H}_4t}\\
&=U_0^{\dag}U_I^{\dag},
\end{split}
\end{equation}
where we have applied the Baker-Campbell-Hausdorff (BCH) and rewritten the sum of the commutators as $\tilde{H}_4$. We have defined $U_0=e^{iH_2t}$ and $U_I=e^{i\tilde{H}_4t}$ in Eq.~\eqref{evovling}. Then the action in the density matrix reads (up to the first order of interaction):
\begin{equation}
\begin{split}
-U^\dag(S_0+S_I)U&=-U_0^\dag U_I^\dag(S_0+S_I)U_IU_0=-U_0^\dag(S_0+S_I+[U_I^\dag,S_0])U_0,
\end{split}
\end{equation}
where $[U_I^\dag,S_0]$ is quartic. Since $U_0$ is a non-interacting evolving operator, the action contains a quadratic term $S_0=-U_0^\dag S_0U_0$ and a quartic term $S_I=-U_0^\dag(S_I+[U_I^\dag,S_0])U_0$, which is similar to the density matrix as discussed in Sec.~\ref{II}. So the density matrix can be written as:
\begin{equation}\label{goalrho}
\rho=e^{-S_0-S_I},
\end{equation}
just as in the thermal equilibrium case. As a consequence, the reduced density matrix in Eq.~\eqref{rhogeneral} can be achieved by expanding $\rho$ around $S_I=0$ and tracing out the Hilbert space of region $\bar{\mathcal{R}}$. Therefore, Eq.~\eqref{res} is valid.

Besides the VNE expression, we have to prove that the results of entanglement measurement still hold in non-equilibrium systems. The key point is the form of correction term of reduced density matrix in the form of Eq.~\eqref{eq.16}. If the interaction is in the bulk, $S_I$ of course contains four CAO of subset $\mathcal{R}$, so the correction term of reduced density matrix is quartic. As for the boundary-interaction case, since there is only one CAO of subset $\mathcal{R}$ in $S_I$, we have to expand the density matrix up to the second order, which is similar as the first order expansion in Eq.~\eqref{evovling}. Since there must be at most two CAO of subset $\mathcal{R}$ in the second order expansion, the correction term in reduced density matrix is quadratic. As a consequence, the results of entanglement measurement through transport still hold: with interaction at the boundary, the correction term in reduced density matrix will be quadratic so Eq.~\eqref{chiedge} is achieved. In contrast, with interaction in the bulk, the correction in reduced density matrix will be quartic so we need $M_{(\mathcal{R}),0}{ii}\to 0$ or $1$ for $\forall i\in\mathcal{H}_{\mathcal{R}}$. Since now we are not in a thermal state, here we need a system initially in a thermal state at nearly zero temperature, i.e., $T(t_i)\to 0$ and the transmission probability is nearly $0$ or $1$, which is a generalization of Klich-Levitov's result~\cite{klich2009quantum}.

\subsubsection{Weakly Fluctuating Case}\label{VB}
To define a saddle point, we have to introduce bosonic field. In $\rho$, we have limited $H_4$ and $S_I$ to be quartic, which allows us to apply H-S transformation. The difference between non-equilibrium case with the thermal equilibrium case is that we need to introduce three bosonic fields in $e^{-i(H_2+H_4)t}$, $e^{-S_0-S_I}$ and $e^{i(H_2+H_4)t}$, respectively:
\begin{equation}
\rho=\int D[{\varphi_1,\varphi_2,\varphi_3}]\frac{e^{-i(H_2+H'_{\varphi_1}+H''_{\varphi_1\varphi_1})t}e^{-S_0-S_{\varphi_3}'-S_{\varphi_3\varphi_3}''}e^{i(H_2+H'_{\varphi_2}+H''_{\varphi_2\varphi_2})t}}{ZZ_{\varphi_1}Z_{\varphi_2}Z_{\varphi_3}},
\end{equation}
where $\varphi_1$, $\varphi_2$, $\varphi_3$ are the introduced bosonic fields, $H_\varphi'~(S'_{\varphi})$ represents the Hamiltonian (action) linear to $\varphi$ and $H''_{\varphi \varphi}~(S''_{\varphi\varphi})$ represents the Hamiltonian (action) linear to the quadratic terms of $\varphi$. The saddle point is defined as $\varphi_0$ where $\partial_\varphi Z_{eff}|_{\varphi_0}=0$ with $Z_{eff}=\Tr_{c}e^{i(H_2+H'_{\varphi_1}+H''_{\varphi_1\varphi_1})t}e^{-S_0-S_{\varphi_3}'-S_{\varphi_3\varphi_3}''}e^{-i(H_2+H'_{\varphi_2}+H''_{\varphi_2\varphi_2})t}$ the effective partition function after tracing out the fermionic Hilbert space, and $\varphi$ a simplified notation of $\varphi_1$, $\varphi_2$, $\varphi_3$. Notably, now the saddle point is where the first partial derivative of all bosonic fields equals $0$. Then if we have a stable saddle point and the saddle-point approximation works, we can expand $\rho$ around the saddle point and keep terms up to the Gaussian fluctuations. After the same steps as in Sec.~\ref{III}, the density matrix reads:
\begin{equation}\label{noneqwf}
\begin{split}
\rho=\frac{e^{-i\tilde{H}_2t}(1+\tilde{G}_1)e^{-\tilde{S}_0}(1-\tilde{G}_3)[e^{-i\tilde{H}_2t}(1+\tilde{G}_1)]^\dag}{Z'},
\end{split}
\end{equation}
with $\tilde{H}_2=H_2+H_{\varphi_{1,0}}'$, $\tilde{S}_0=S_0+S_{\varphi_{3,0}}'$ where $H_{\varphi_{1,0}}'$ and $S_{\varphi_{3,0}}'$ are the saddle-point contributions. $\tilde{G}$ are the quartic correction arising from Gaussian fluctuations, just as in Eq.~\eqref{rhof}. If we define $U_0'^\dag=e^{-i\tilde{H}_2t}$ and $U_I'^\dag=(1+\tilde{G}_1)$, Eq.~\eqref{noneqwf} can be found similar to Eq.~\eqref{noneqwp}, so we can apply the same proof as in Sec.~\ref{VA}. As a result, the density matrix can be written in the form of Eq.~\eqref{goalrho}, which leads to the validity of Eq.~\eqref{vnesaddle}. Besides, the number of CAO of the correction term can be analyzed as in Sec.~\ref{IV}: $\tilde{G}_{1,3}$ contains four CAO of subset $\mathcal{R}$ with interaction in the bulk. As for the boundary-interaction case, $\tilde{G}$ generally contains two CAO. As a consequence, Eqs.~\eqref{chibulk} and~\eqref{chiedge} in Sec.~\ref{IV} still hold.

\bibliography{ref.bib}

\end{document}